\documentclass[twocolumn,epj,nopacs]{svjour}

\usepackage{atlasphysics}
\usepackage{graphicx}
\usepackage{xspace}
\usepackage{units}
\usepackage{textpos}
\usepackage{amssymb}
\usepackage{cite,mcite}
\usepackage{hyperref}

\newcommand{\CL}{CL\xspace}

\renewcommand{\MET}{\ensuremath{E_{\mathrm{T}}^{\mathrm{miss}}}\xspace}

\newcommand{\Fig}{Fig.\xspace}
\newcommand{\Ref}{Ref.\xspace}
\newcommand{\Tab}{Table\xspace}

\renewcommand{\mass}[1]{\ensuremath{m_{#1}}}
\renewcommand{\ET}{\ensuremath{E_{\mathrm{T}}}\xspace}
\renewcommand{\antibar}[1]{\ensuremath{#1\bar{#1}}}
\renewcommand{\ttbar}{\antibar{t}\xspace}
\newcommand{\myW}{\ensuremath{W(\to e\nu)+\mathrm{jets}/\gamma}\xspace}
\newcommand{\myttbar}{\ensuremath{\ttbar(\to e\nu)+\mathrm{jets}}\xspace}

\newcommand{\QCDg}{\ensuremath{\mathrm{QCD}_{\gamma}}\xspace}

\newcommand{\neutralino}{\ensuremath{\tilde{\chi}^{0}_{1}}\xspace}

\newcommand{\gravitino}{\ensuremath{\tilde{G}}\xspace}
\newcommand{\Pythia}{{\tt PYTHIA}\xspace}
\newcommand{\Suspect}{{\tt SUSPECT}\xspace}
\newcommand{\Sdecay}{{\tt SDECAY}\xspace}
\newcommand{\Prospino}{{\tt PROSPINO}\xspace}
\newcommand{\btheta}{\bf \theta}

\newcommand{\integLumi}{\unit[36]{\ipb}\xspace}
\newcommand{\integLumiE}{\unit[$\left(36\pm 1\right)$]{\ipb}\xspace}
\newcommand{\Nlimit}{\ensuremath{3.0}\xspace}
\newcommand{\UEDlimit}{\ensuremath{1/R < \unit[961]{GeV}}\xspace}
\newcommand{\UEDlimitlow}{\ensuremath{1/R > \unit[961]{GeV}}\xspace}
\newcommand{\UEDlimitExp}{\ensuremath{1/R < \unit[961]{GeV}}\xspace}

\newcommand{\GGMlimit}{\unit[560]{GeV}\xspace}
\newcommand{\GGMlimitExp}{\unit[560]{GeV}\xspace}
\newcommand{\Nsignal}{762\xspace}
\newcommand{\nbackground}{\ensuremath{0.10 \pm 0.04 (\mathrm{stat}) \pm 0.05 (\mathrm{syst})}\xspace}
\newcommand{\sigmaGGM}{\ensuremath{\sigma < \unit[0.38-0.65]{pb}}\xspace}
\newcommand{\sigmaUED}{\ensuremath{\sigma < \unit[0.18-0.23]{pb}}\xspace}

\begin{document}


\title{Search for Diphoton Events with Large Missing Transverse Energy
with \integLumi of \unit[7]{TeV} Proton-Proton Collision Data with the ATLAS Detector
\begin{textblock}{6}[0,1](0,-3){\small CERN-PH-EP-2011-091; Submitted to EPJC Letters.} \end{textblock}
}
\titlerunning{Search for Diphoton Events with \MET with the ATLAS Detector}

\author{The ATLAS Collaboration}
\institute{CERN, 1211 Geneva 23, Switzerland \email{atlas.publications@cern.ch}}

\date{}

\abstract{
  Making use of \integLumi of proton-proton collision data at $\sqrt{s} =
  \unit[7]{TeV}$, the ATLAS Collaboration has performed a search for
  diphoton events with large missing transverse energy. Observing no
  excess of events above the Standard Model prediction, a
  \unit[95]{\%}~Confidence Level (\CL) upper limit is set
  on the cross section for new physics of \sigmaGGM in the context of a generalised
  model of gauge mediated supersymmetry breaking (GGM) with a
  bino-like lightest neutralino, and of \sigmaUED in the context of a specific model with
  one universal extra dimension (UED). A \unit[95]{\%}~\CL lower limit of \GGMlimit, for
  bino masses above \unit[50]{GeV}, is set on the GGM gluino mass, while a lower limit of \UEDlimitlow
  is set on the UED compactification radius $R$. These limits provide
  the most stringent tests of these models to date.
}

\maketitle

\section{Introduction}
In high-energy proton-proton~($pp$) collisions, Standard Mo\-del (SM)
production of diphoton~($\gamma \gamma$) events with two prompt
photons and large missing transverse energy (\met) is due primarily to
$W/Z$ production with the associated electroweak production of two
photons.
Taking into account the branching
ratios of $W/Z$ decays that include at least one neutrino, the cross sections
for events with large transverse energy~(\ET) photons are
well below \unit[100]{fb} for \unit[7]{TeV} $pp$ collisions. In contrast, some new physics models
predict much larger $\gamma \gamma + \met$ yields.
This Letter reports on the search for $\gamma \gamma + \met$ events
with \integLumi of ATLAS data collected in 2010, extending a prior
study~\cite{Aad:2010qr} performed with \unit[3.1]{\ipb}.
The results are interpreted in the context of a general model of
gauge-mediated supersymmetry
breaking (GGM)~\cite{Meade:2008wd,*Buican:2008ws,Ruderman:2011vv} as well as a model
positing one universal extra
dimension (UED)~\cite{Macesanu:2002ew,*Macesanu:2005jx}.

\section{Supersymmetry}
Supersymmetry~(SUSY)~\cite{Golfand:1971iw,*Neveu:1971rx,*Neveu:1971iv,*Ramond:1971gb,*Volkov:1973ix,*Wess:1973kz,*Wess:1974tw} 
introduces a symmetry between fermions and bosons,
resulting in a SUSY partner with identical quantum numbers (differing by half a unit of spin)
for each SM particle.
At the Large Hadron Collider~(LHC), the dominant SUSY process would be the production of
pairs of SUSY partners of quarks (squarks) or gluons (gluinos) via the strong interaction.  These would then decay through
cascades involving other sparticles until the lightest SUSY particle (LSP) is produced.
Assuming $R$-parity conservation~\cite{Fayet:1977yc,*Farrar:1978xj}, the LSP is stable and
escapes detection.

No SUSY partners of SM particles have been observed yet, indicating that SUSY must be
broken to decouple the masses of the SM particles and their SUSY partners.
In gauge-mediated SUSY breaking~(GMSB)
models~\cite{AlvarezGaume:1981wy,*Dine:1981,*Dimopoulus:1981,*Nappi:1982,*Dine:1993yw,*Dine:1994vc,*Dine:1995ag}
the LSP is the gravitino \gravitino. GMSB experimental signatures are
largely determined by the character of the next-to-lightest SUSY
particle~(NLSP), which for most of the GMSB parameter space is the lightest neutralino
\neutralino. Should this neutralino be the SUSY partner of the U(1)
gauge boson (the ``bino''), the final decay in the cascade is dominated
by $\neutralino\to\gamma\gravitino$, with two cascades
per event, leading to a final state
$\gamma\gamma+\met+X$, where $\met$ results from the escaping
gravitinos and $X$ represents SM particles emitted in the prompt
cascade decays.

Previous searches for GMSB~\cite{Aaltonen:2009tp,Abazov:2010us} were
performed using a minimal GMSB model~\cite{Allanach:2002nj}. To reduce
the number of free parameters in this model, several assumptions are
made, including gaugino mass unification. These assumptions lead to a
mass hierarchy in which squarks and gluinos are much heavier than the
lightest neutralino and chargino. For example, the current lower
limit on the minimal GMSB neutralino mass of \unit[175]{GeV}~\cite{Abazov:2010us},
based on the assumption of direct production of charginos and neutralinos and obtained for 
the large integrated luminosity accumulated at the Tevatron,
corresponds to gluino and squark mass limits in the TeV range.
The data analysed in this Letter have a limited sensitivity to the
minimal GMSB model and do not improve upon the current Tevatron limits.

In the following a GGM SUSY
model~\cite{Meade:2008wd,*Buican:2008ws,Ruderman:2011vv} is
considered for which the
gluino and neutralino masses are treated as free parameters. 
The other sparticle masses were fixed to $\sim\unit[1.5]{TeV}$,
leading to a dominant production mode at $\sqrt{s} = \unit[7]{TeV}$ of a pair of
gluinos that would decay via cascades into the neutralino NLSP.
The NLSP was taken to be
the bino-like neutralino, decaying promptly ($c\tau_{\mathrm{NLSP}} <
\unit[0.1]{mm}$) into a photon and a gravitino. The branching
fraction to $\gamma\gravitino$ is almost \unit[100]{\%} for low
neutralino masses and approaches $\cos^2\theta_W$ for neutralino
masses well above the $Z$ mass.
The Higgsino mass term, $\mu$, was set to \unit[1.5]{TeV} and the ratio of the two Higgs vacuum
expectation values, $\tan\beta$, to $2$. 
Similar parameter values were used in the only previous study of this
model, in which the CMS experiment excluded signal cross sections
between 0.3 and \unit[1.1]{pb} at \unit[95]{\%} Confidence Level~(\CL) and determined
limits on the squark, gluino and neutralino
masses~\cite{Chatrchyan:2011wc}.
These limits derive from the copious production expected for
accessible colour-charged sparticles (squarks, gluinos), and benefit
from the increased collision energy even for relatively low
integrated luminosity, extending the reach of
the LHC beyond that of the current Tevatron
data~\cite{Ruderman:2011vv}.

\section{Extra dimensions}
UED models~\cite{Appelquist:2000nn} postulate the existence of additional 
spatial dimensions in which all SM particles can propagate, leading 
to the existence of a series of excitations for each SM particle
known as a Kaluza-Klein (KK) tower. This analysis considers
the case of a single UED, with compactification radius 
$R \approx \unit[1]{TeV^{-1}}$.
The masses of the states of successive levels in the tower are separated by $\approx 1/R$;
for a given KK level, the approximate mass degeneracy of the KK excitations is
broken by radiative corrections~\cite{Cheng:2002ab}.  
The lightest KK particle (LKP) is the KK photon of the first level, denoted $\gamma^*$.
At the LHC, the main UED process would be the production via the strong interaction
of a pair of first-level KK quarks and/or gluons~\cite{Macesanu:2002db}, which would
decay via cascades involving other KK particles until reaching the LKP at the end of the decay chain.
If the UED model is embedded in a larger space with
$N$ additional eV$^{-1}$-sized dimensions accessible 
only to gravity~\cite{DeRujula:2000he}, the LKP could decay gravitationally via
$\gamma ^* \rightarrow \gamma + G$~\cite{Macesanu:2002ew,*Macesanu:2005jx}, where $G$ represents one 
of a tower of eV-spaced graviton states, leading to a distribution of graviton mass between $0$ and $1/R$. 
With two decay chains per event, the final state
would again be $\gamma \gamma + \met + X$, where \met \  results from
the escaping gravitons and  $X$ represents SM particles emitted in the cascade decays.

The UED model considered here is defined by specifying $R$ and $\Lambda$,
the ultraviolet cut-off used in the calculation of radiative
corrections to the KK masses. This analysis sets $\Lambda $ such that
$\Lambda R = 20$~\cite{Cheng:2002ab}.  For $1/R = \unit[700]{GeV}$, the 
masses of the first-level KK photon, quark and gluon are $700$, $815$
and $865$~GeV, respectively~\cite{ElKacimi:2009zj}.  The $\gamma ^*$
mass is insensitive to $\Lambda $, while other KK masses typically change by
a few percent when varying $\Lambda R$ in the range $10-30$.
The gravitational decay widths of the KK particles are set by $N$ and
$M_D$, the Planck scale in the ($4+N$)-dimensional theory.  For the
chosen values of $N = 6$ and $M_D = \unit[5]{TeV}$, and provided $1/R \lesssim
\unit[1]{TeV}$, the LKP is the only KK particle to have an appreciable rate of
gravitational decay.
The same parameter values were used in prior searches by D0 and ATLAS.
D0 excluded values of $1/R < \unit[477]{GeV}$ at \unit[95]{\%}~\CL~\cite{Abazov:2010us}.
ATLAS analysed \unit[3.1]{\ipb} of $pp$ collisions data at $\sqrt{s} =
\unit[7]{TeV}$, excluding values of $1/R < \unit[729]{GeV}$ at
\unit[95]{\%}~\CL~\cite{Aad:2010qr}.
 
\section{Simulated samples}
For the GGM model,
the full mass spectrum and the gluino branching ratios and
decay widths were calculated
for a range of gluino and neutralino masses 
using \Suspect 2.41~\cite{Djouadi:2002ze}
and \Sdecay~1.3~\cite{Muhlleitner:2003vg}. 
The SUSY Les Houches Accord files for the model used may be found
online~\cite{slhaonline}.
The Monte Carlo~(MC) signal samples were produced using
\Pythia~6.423~\cite{pythia} with {\tt MRST2007
  LO${}^{*}$}~\cite{Sherstnev:2007nd} parton distribution
functions~(PDF).  
Cross sections were calculated at next-to-leading order (NLO) using
\Prospino 2.1~\cite{Beenakker:1996ch}.
In the case of the UED model, a range of $1/R$ values was generated using the
leading-order (LO) implementation of the UED model in 
\Pythia~\cite{ElKacimi:2009zj}; no NLO calculations of UED processes
are currently available.
For both sets of samples, the MC10 parameter tune~\cite{mc10AMBT1} was
used. All signal and SM background samples were passed through a {\tt
  GEANT4}-based simulation~\cite{geant4} of the ATLAS detector~\cite{Aad:2010wq}
including the simulation of pileup (multiple $pp$ collisions within the
same bunch crossing) of approximately 2 events per bunch crossing,
similar to that seen in the data. The MC samples were reconstructed
with the same algorithms that were used for the data.

\section{ATLAS detector}
The ATLAS detector~\cite{Aad:2008zzm} is a multi-purpose 
apparatus with a forward-backward symmetric cylindrical
geometry and nearly 4$\pi$ solid angle coverage.
Closest to the beamline are tracking detectors that use layers of
silicon-based ($\left|\eta\right|<2.5$) and straw-tube ($\left|\eta\right|<2.0$) detectors~\footnote{ATLAS uses a right-handed coordinate system with its origin at the nominal interaction point (IP) in the centre of the detector and the $z$-axis along the beam pipe. The $x$-axis points from the IP to the centre of the LHC ring, and the $y$ axis points upward. Cylindrical coordinates $(R,\phi)$ are used in the transverse plane, $\phi$ being the azimuthal angle around the beam pipe. The pseudorapidity is defined in terms of the polar angle $\theta$ as $\eta=-\ln\tan(\theta/2)$.},  located inside a thin superconducting solenoid 
that provides a 2~T magnetic field, to measure the trajectories of
charged particles. The solenoid is surrounded by
a hermetic calorimeter system.  A lead liquid-argon (LAr) sampling calorimeter is divided into
a central barrel calorimeter and two end-cap calorimeters, each housed in a
separate cryostat. Fine-granularity LAr electromagnetic (EM)
calorimeters provide coverage for $\left|\eta\right| < 3.2$ to precisely measure
the energy and position of electrons and photons.
In the region $\left|\eta\right| < 2.5$, the EM calorimeters are segmented into three
longitudinal layers.  The second layer, in which 
most of the EM shower energy is deposited,
is divided into cells of  granularity of $\Delta \eta \times
\Delta \phi = 0.025 \times 0.025$, 
with the first layer segmented with finer granularity~\cite{Aad:2008zzm} to provide
rejection of neutral mesons in jets when selecting isolated photons.
A presampler, covering $\left|\eta\right| < 1.8$,  is used to correct for
energy lost upstream of the calorimeter. An iron-scintillator
tile calorimeter provides hadronic coverage in the range 
$\left|\eta\right| < 1.7$. In the end-caps
($\left|\eta\right| > 1.5$), LAr hadronic calorimeters match the outer $\left|\eta\right|$
limits of the end-cap EM calorimeters.
LAr forward calorimeters provide both EM
and hadronic energy measurements, and extend the coverage to
$\left|\eta\right| < 4.9$, which is required for reliable \MET measurements.
An extensive muon spectrometer system that incorporates large
superconducting toroidal magnets lies outside the calorimeter system.

\section{Photon and \MET reconstruction}
The reconstruction of photons is described in detail in \Ref~\cite{Aad:2010sp}. To
select photon candidates, EM calorimeter clusters were required to satisfy several quality criteria and
lie outside problematic calorimeter regions. 
EM calorimeter clusters without any matching track
were considered to be unconverted photons; clusters that
have a track match were considered to be electrons; clusters matched to a
conversion vertex were considered to be converted photons. A conversion
vertex is either a vertex that has two electron-like tracks, as
determined by transition radiation in the straw-tube tracker, and is consistent
with a massless parent object, or one electron-like track that
has no track hits in the pixel vertexing layer. 
Photon candidates were required to be within $\left|\eta\right| < 1.81$ and to
be outside the transition region $1.37 < \left|\eta\right| < 1.52$
between the barrel and the end-cap calorimeters. 
The analysis used ``loose'' and ``tight'' photon
selections~\cite{Aad:2010sp}, which include a limit on the fraction of the energy in the hadronic calorimeter as well as 
requirements that the transverse width of the shower, measured in the
middle layer of the EM calorimeter, be consistent with the narrow width expected for an EM shower.  
The tight photon selection uses, in addition, shape information from the first sampling layer to distinguish between isolated photons and photons from the decay of neutral mesons. 
The tight selection is approximately \unit[85]{\%} efficient for both unconverted
and converted photons while rejecting most of the background from hadronic jets~\cite{photonperf}.

The reconstruction of \MET was based on topological calorimeter clusters~\cite{Aad:2010yt} with $|\eta | < 4.5$.
The cluster energy was calibrated to correct for the non-compensating calorimeter response, energy losses in 
dead material, and out-of-cluster energies. 
Rare events with large transverse energies that are
unrelated to the collision, concentrated in a few cells, and due mainly to 
electronic discharges and noise, have been observed.
Cuts were applied to eliminate such backgrounds,
rejecting fewer than 0.3\% of the selected events while having a
negligible impact on the GGM and UED signal efficiencies.

\section{Data analysis}
The data sample was collected during stable beam periods of
\unit[7]{TeV} $pp$ collisions at the LHC, and corresponds to an
integrated luminosity of \integLumiE.  Selected events had to satisfy a
trigger requiring either a single EM cluster with $\ET > \unit[14]{GeV}$
during the first \unit[0.5]{\ipb} of $pp$ collisions or two loose photon candidates with
$\ET > \unit[15]{GeV}$ thereafter.
Furthermore, they had to contain at least one reconstructed primary vertex
consistent with the average beam spot position and with at least five associated tracks.
The trigger and vertex requirements are very close to \unit[100]{\%} efficient for signal MC events.

Events were retained if they had at least two tight photon candidates, one with $\ET > \unit[30]{GeV}$
and the other with $\ET > \unit[20]{GeV}$. In addition, a photon
isolation cut was applied, whereby the \ET in a cone of radius $0.2$ in the
$\eta$-$\phi$ space around the centre of the cluster, excluding all the
cells belonging to the cluster (in a region with a size corresponding to $5\times7$
cells in $\eta\times\phi$ in the second layer of the EM calorimeter),
had to be less than \unit[10]{\%} of the transverse energy of the cluster.
This requirement has a signal efficiency greater than \unit[93]{\%} 
while rejecting a significant portion of the background from multijet events.

The signal region was defined as $\MET>\unit[125]{GeV}$, chosen to
maximise sensitivity to the GGM and UED signal final states.

\begin{figure}[tb]
  \begin{center}
    \includegraphics[width=0.48\textwidth]{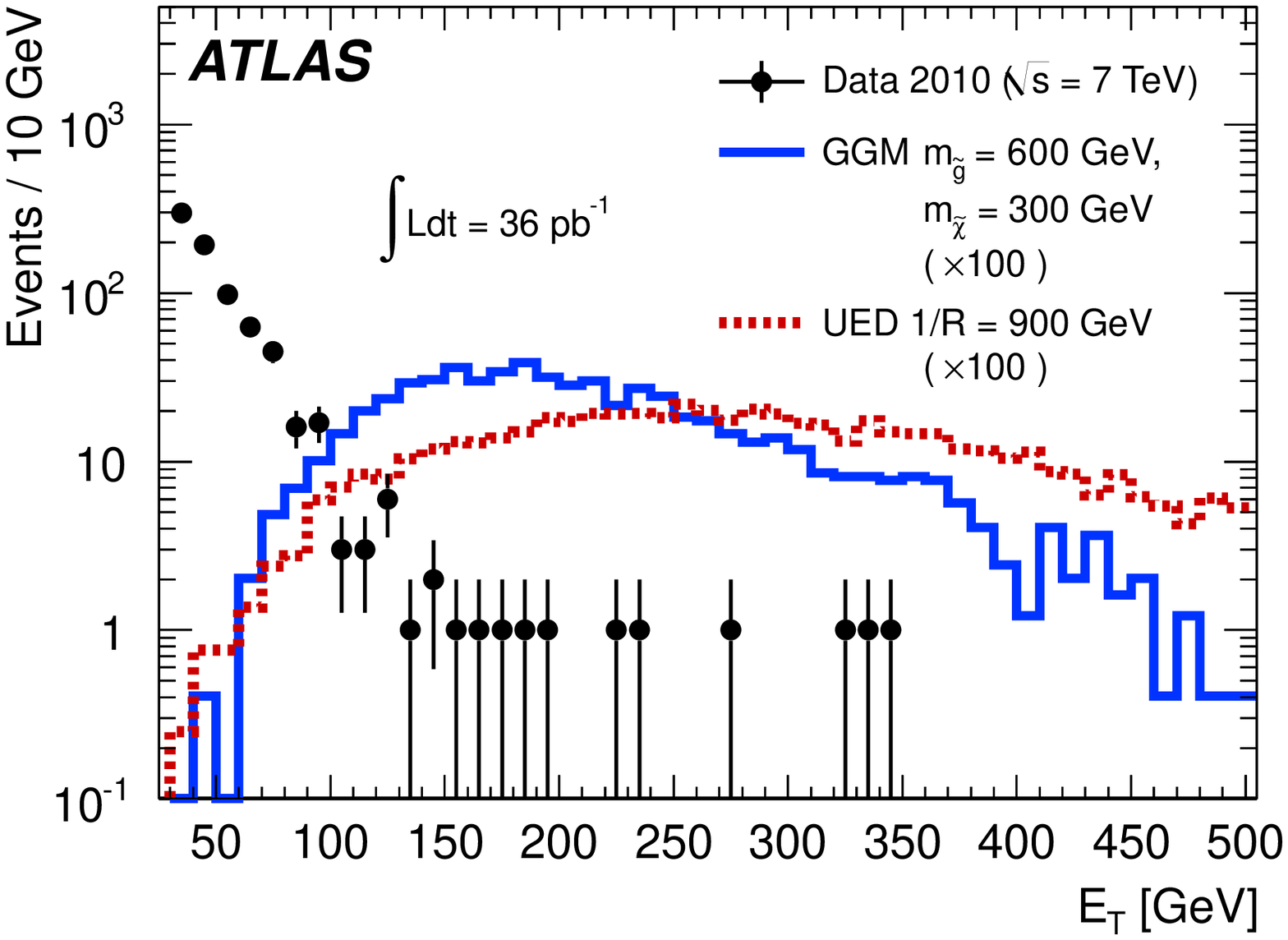}
  \end{center}
  \caption{The \ET spectrum of the leading photon in the $\gamma\gamma$
    candidate events (points, statistical uncertainty only) together with the spectra from simulated GGM
    ($\mass{\gluino}/\mass{\neutralino} = \unit[600/300]{GeV}$) and UED
    ($1/R = \unit[900]{GeV}$) samples, prior to the application of the 
    $\MET>\unit[125]{GeV}$ cut.
    \label{fig:PhoPt}
  }
\end{figure}

A total of \Nsignal $\gamma\gamma$ candidate events were observed
passing all selection cuts except the \MET cut. 
After the $\MET > \unit[125]{GeV}$ cut, no candidate events survive in the
data.
The \ET of the leading
photon for events in this sample is shown in \Fig~\ref{fig:PhoPt}. Also shown is 
the \ET spectrum obtained from GGM MC samples for
$\mass{\gluino}=\unit[600]{GeV}$ and
$\mass{\neutralino}=\unit[300]{GeV}$ and from UED MC samples for $1/R =
\unit[900]{GeV}$, representing model parameters near the expected
exclusion limit.

\section{Background estimation}
Guided by the procedure developed in \Ref~\cite{Abazov:2010us}, 
the number of large~\MET diphoton events from SM sources 
can be grouped into two primary components
and estimated with dedicated control samples.
The first of these components, referred to as QCD background,
arises from a mixture of SM processes that include $\gamma \gamma$ production as well as
$\gamma $ + jet and multijet events with at least one jet misidentified as a photon.
The second background component is due to $W+X$ and \ttbar events, 
for which final-state neutrinos produce significant \met.
These can pass the selection if an electron from the
$W$ or $t$-quark decay is misidentified as a
photon and the second photon is either a real photon
($W\gamma$ events), a jet faking a photon ($W$ + jets events),
or a jet or second electron faking a photon (\ttbar events).

In order to estimate the QCD background from $\gamma$ + jet and
multijet events, an independent ``\QCDg'' control sample, designed to
provide a model of the \met \ response for events with jets faking
photons, was defined by selecting
events for which at least one of the photon candidates did not pass the tight photon identification. 
The background from QCD events producing two prompt photons
was modeled using the \MET spectrum measured in a 
high-purity sample of \Zee \ events, with no additional jets, selected
by requiring two electrons~\cite{Aad:2010yt} with $\ET>\unit[30]{GeV}$ and $\ET>\unit[20]{GeV}$, respectively.
Both electrons are required to have $\left|\eta\right|<2.47$, excluding the transition region $1.37 <
\left|\eta\right| < 1.52$.
In addition, the dielectron invariant mass was required to be consistent with the $Z$ mass.
As confirmed by MC simulation, the \MET spectrum of the $\Zee$ sample with no additional jets,
which is dominated by the calorimeter response to two genuine EM
objects, accurately represents the \MET response of SM $\gamma\gamma$
events.

The  QCD background is the dominant source of observed $\gamma \gamma$ events at low \MET and its spectrum, which contains a mixture of events with zero, one or two prompt photons, is expected to lie between the spectra from the \QCDg and \Zee \ control samples.
The \met \ spectrum of the \QCDg control sample,
which provides the best description of the \MET spectrum
at low \met,  was chosen to model the composite QCD background. 
The difference between this estimate and that derived from the \Zee \ template 
was used to provide an estimate of the systematic uncertainty
on the resulting background prediction.
The QCD background was normalised to have
the same number of events  as the $\gamma \gamma$ candidate sample in the
region $\MET<\unit[20]{GeV}$, where
contributions from events with
genuine \MET, such as $W+X$ and $\ttbar$ events, can be neglected.  
It should be noted that a possible background contribution from $Z+X$ events,
with the $Z$ boson decaying to neutrinos, would be incorporated within
this estimate of the QCD background, since it would enter the
signal region through the misidentification of jets as photons.

\begin{figure}[tb]
\begin{center}
\includegraphics[width=0.48\textwidth]{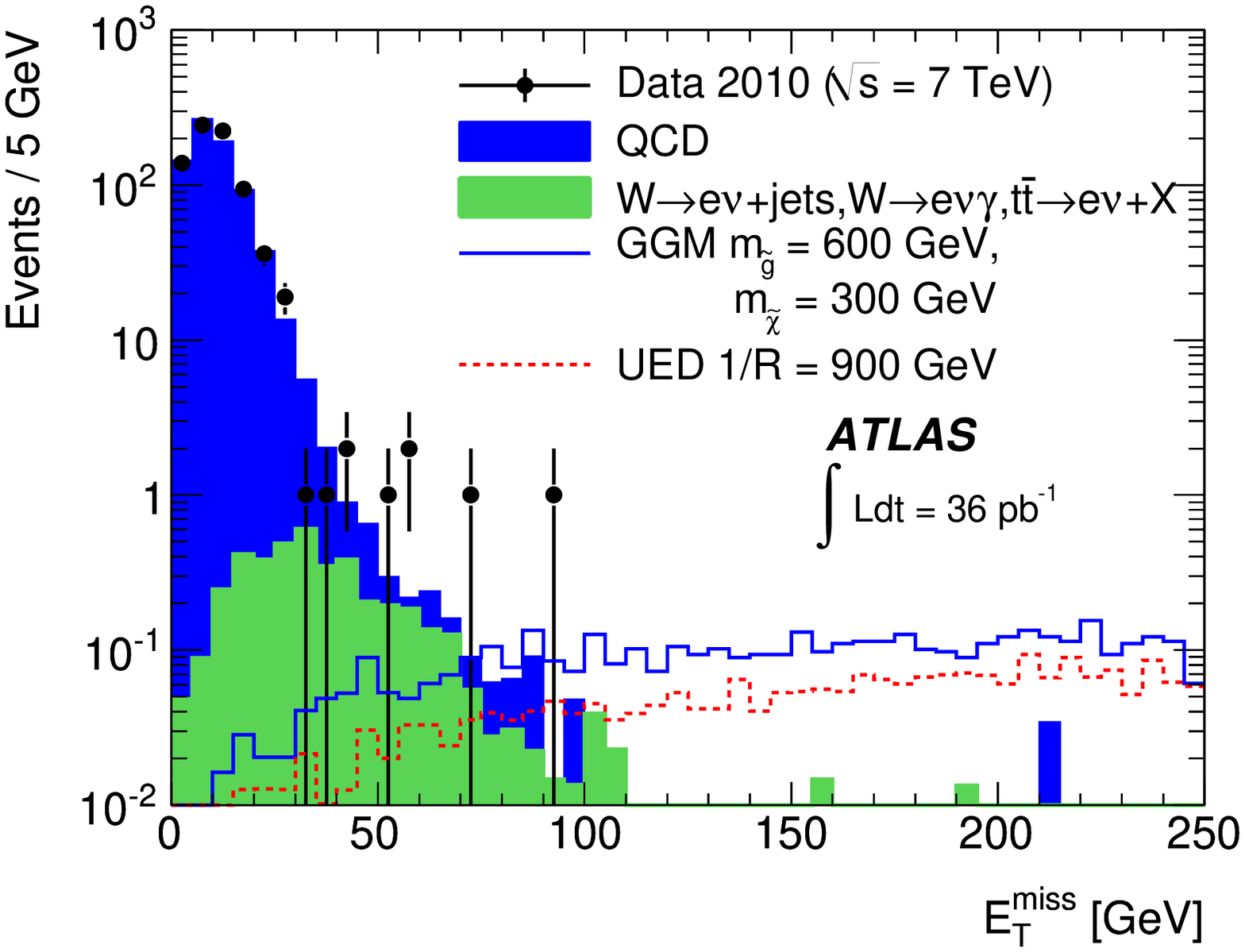}
\end{center}
\caption{\MET spectra of the $\gamma\gamma$ candidates (points, statistical uncertainty only) and estimated
  background from the QCD (normalised to the number of $\gamma\gamma$
  candidates with $\MET < \unit[20]{GeV}$) and $\myW$ and $\myttbar$
  sources, together with the spectra from simulated GGM
  ($\mass{\gluino}/\mass{\neutralino} = \unit[600/300]{GeV}$) and UED
  ($1/R = \unit[900]{GeV}$) samples. }
\label{fig:bkgnd}
\end{figure}

The \QCDg template has one event with 
\MET greater than \unit[125]{GeV}, whereas the \Zee \ template has none.
Taking into account the expected ratio of events in the portion of the control 
region with $\MET > \unit[125]{GeV}$ to those in the signal region, this
leads to a QCD background prediction of $0.034\pm0.034(\mathrm{stat})\pm0.034(\mathrm{syst})$ events.  
The \QCDg \met \ spectrum is shown together with 
the $\gamma\gamma$ sample in \Fig~\ref{fig:bkgnd}.  

\begin{figure}[t]
\begin{center}
\includegraphics[width=0.48\textwidth]{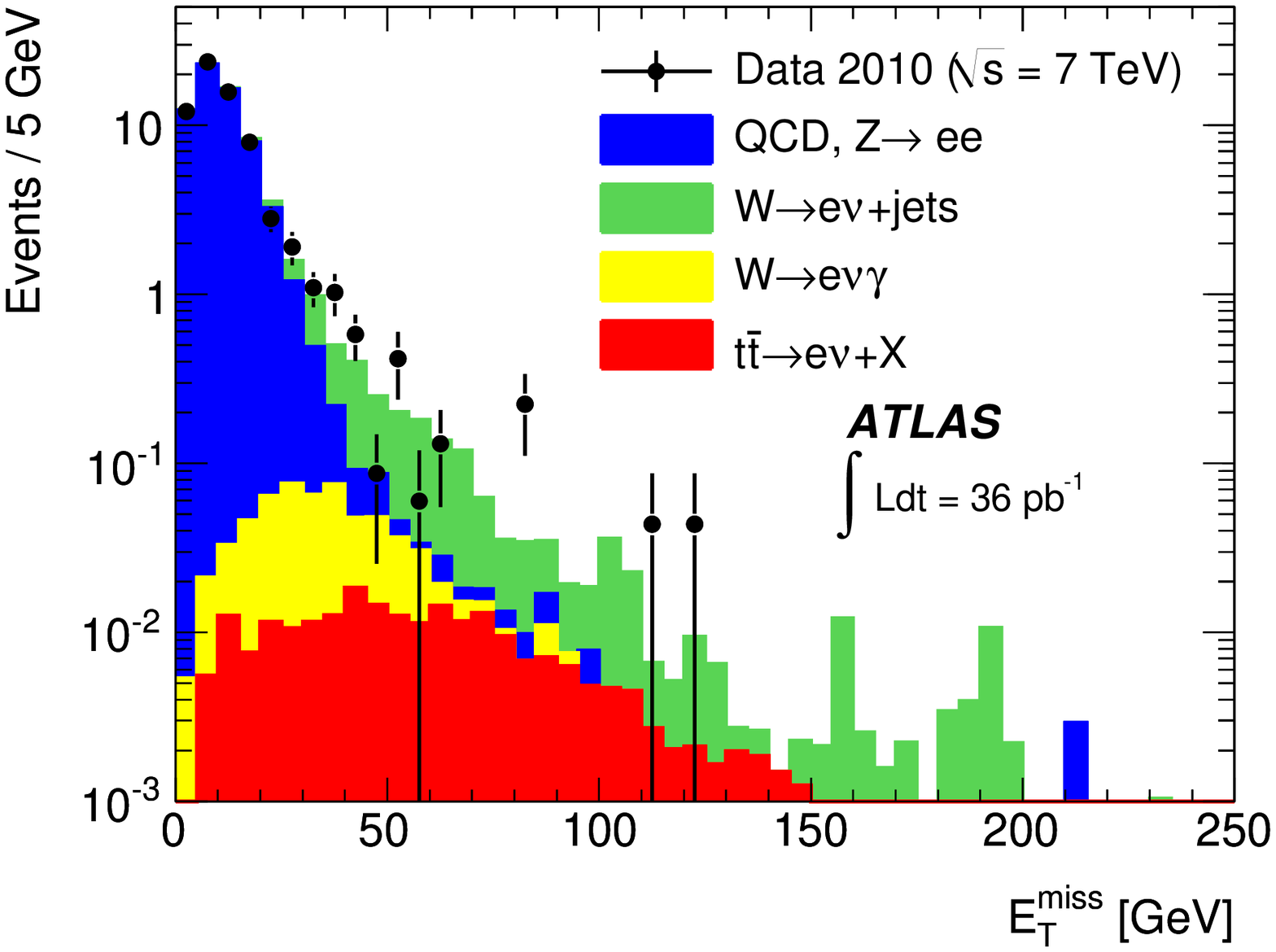}
\end{center}
\caption{\MET spectrum of the electron-photon control sample (points, statistical uncertainty only), normalised according to the probability for an  
electron to be misidentified as a tight photon, compared to the sum of the expected
background, broken down by component (stacked histograms).
For the purpose of this comparison, the 
expected contribution from $\myW$ and \myttbar events is taken directly from 
MC. At low values of \MET, the spectrum is dominated by QCD events and 
\Zee \ events for which one of the electrons has been misidentified as
a photon. These two classes of events are accounted for (blue histogram) by scaling
the \QCDg template to match the interval of the spectrum in the
region $\MET < \unit[20]{GeV}$.}

\label{fig:ElPho}
\end{figure}

The second significant background contribution,
from $W+X$ and \ttbar events, was estimated
via an ``electron-photon'' control sample composed of events with 
both a photon and an electron
with $\ET>\unit[20]{GeV}$, with the additional requirement that
either the electron or photon has $\ET>\unit[30]{GeV}$, and scaled by the probability for an
electron to be misidentified as a tight photon.  This probability was
estimated by comparing the rate at which photons and
electrons pair with electrons to form an invariant mass consistent with that of the $Z$ boson.
The misidentification probability with the selection cuts used in this
analysis varies between \unit[5]{\%} and \unit[12]{\%} as a function
of $\eta$, since it depends on the amount of material in the inner
detector.
The \MET spectrum for this control sample is shown in \Fig~\ref{fig:ElPho}, compared
to the expected contributions from various background sources.  The electron-photon control sample has a 
significant contribution from \Zee \ events, for which one electron fakes a photon, and from QCD. 
Both of these contributions 
must be subtracted in order to predict the contribution to the \MET 
distribution from events with genuine \MET, such as $W+X$ and $t\bar{t}$ events. 
The contribution from QCD and \Zee \ events was estimated
by normalizing the \QCDg \MET distribution
to the scaled electron-photon \MET distribution in the region $\MET < \unit[20]{GeV}$, as shown in \Fig~\ref{fig:ElPho}.
This distribution was then subtracted from the 
scaled electron-photon control sample, 
yielding a prediction for the contribution to the high-\MET 
diphoton sample from $W+X$ and \ttbar events.  
For $\MET > \unit[30]{GeV}$, the sample is dominated by events with genuine \MET and electrons from $W\to e\nu$.

No events with $\MET > \unit[125]{GeV}$ were observed in the electron-photon control sample;
taking into account the measured electron-to-photon misidentification rate, this yields an 
upper limit on the background from $W+X$ and \ttbar of $0.093$ events at \unit[90]{\%}~\CL.
Two additional methods were employed to increase the statistical power of the electron-photon
control sample. First, the photon selection requirement was loosened for the control sample, leading to an observation 
of one event in the signal region, and a corresponding background estimate of $0.027 \pm 0.027(\mathrm{stat})$
events. Second, MC samples of the three main components ($W+\mathrm{jets}$, $W+\gamma$ and \ttbar) were used to model the region where data is limited.  
The MC samples, which had the same electron and photon selection and scaling applied as for the electron-photon control sample, were normalised to the control sample, after the QCD and \Zee \ subtraction, in the range $\MET > \unit[40]{GeV}$ (to avoid the QCD and \Zee \ dominated region at low \MET). 
This MC extrapolation was used to determine the background with 
$\MET > \unit[125]{GeV}$, and led to an estimate of $0.057 \pm 0.015(\mathrm{stat})$ background 
events. 
The estimate from the loosened-photon control sample
provides a lower bound on the systematic uncertainty of this background
component that lies $0.030$ events below this central value. In
addition, a gaussian distribution with a mean of $0.057$ and variance
of $0.031$ events has \unit[90]{\%} of its probability density below
$0.093$ events, providing an upper bound of $0.031$ events above the
central value. Thus, a value of $\pm 0.031$ events was chosen for the systematic uncertainty,
yielding an estimate of $0.057\pm0.015(\mathrm{stat})\pm0.031(\mathrm{syst})$ for the background from $W+X$ and \ttbar events. 
Also included in the quoted systematic uncertainty is the uncertainty ($\pm \unit[10]{\%}$) on the electron-photon 
fake rate and the uncertainty ($\pm \unit[6]{\%}$) on the subtraction of the QCD contribution to the
scaled electron-photon control sample derived by replacing the \QCDg distribution with that of the
$\Zee$ sample when performing the subtraction.

In addition to the QCD, $W+X$ and \ttbar contributions, a small irreducible background contribution
of $0.004$ events from $W(\to e\nu)\gamma\gamma$ and $Z(\rightarrow\nu\nu)\gamma\gamma$ events was estimated
directly from MC simulation. Backgrounds due to cosmic-ray and non-collision activity were suppressed through the use of timing and jet-quality cuts and by
the primary vertex requirement, and are not expected to contribute singificantly in
the signal region.

Figure~\ref{fig:bkgnd} shows the \met \ spectrum of selected $\gamma \gamma$
candidates, superimposed on the total background prediction.
Table~\ref{tab:events} summarises the number of observed $\gamma \gamma$ candidates, as
well as the expected backgrounds and two representative
GGM and UED signal contributions, in several \met \ ranges. 
No indication of an excess at high \met \ values, where UED and
GGM are expected to dominate, is observed.
No events are observed in the signal region $\MET > \unit[125]{GeV}$, 
compared to an expectation of \nbackground background events.

\begin{table*}[hbpt]
\caption{Numbers of observed $\gamma \gamma$ candidates, SM background candidates estimated from the \QCDg \ and 
electron-photon control samples and $\myW$ and \myttbar MC,
and expected signal events from GGM with $(\mass{\gluino}/\mass{\neutralino}) = 
\unit[(600/300)]{GeV}$ and UED with $1/R = \unit[900]{GeV}$, 
given in various \met \ ranges. The uncertainties are statistical only.
 The first row, for $\met < \unit[20]{GeV}$, is the control region used to normalise the 
QCD background  to the number of observed $\gamma \gamma$ candidates.  The total predicted background events also include the small contribution from $W(\rightarrow e \nu)\gamma\gamma$ and $Z (\rightarrow \nu\nu)\gamma\gamma$ events.}
\label{tab:events}
\center
\begin{tabular}{c|r|ccc|cc} 
\hline
\met \ range  & \multicolumn{1}{c|}{Data}  & \multicolumn{3}{c|}{Predicted background events} & \multicolumn{2}{c}{Expected signal events} \\
   $[$GeV$]$  & events & Total & QCD & $W/\ttbar(\rightarrow e \nu)+\mathrm{jets}/\gamma$  & GGM & UED \\ \hline

  \hspace{0.06in}0 - 20  & 698 & \hspace{0.02in}- & \hspace{0.06in}-  &            -    & $0.05\pm 0.01$ & $0.02\pm 0.01$  \\ \hline
20 - 75 & 63  & $61.4 \pm 2.3\hspace{0.06in}$ &  $58.3 \pm 2.2$ & $2.99 \pm 0.12$  & $0.59  \pm 0.05$ & $0.25 \pm 0.02$ \\
\hspace{0.06in}75 - 125   &   1   &  $\hspace{0.00in}0.38 \pm 0.08$ & $\hspace{0.06in}0.17 \pm 0.08$ & $0.19\pm 0.03$ & $0.96 \pm 0.06$ & $0.43 \pm 0.02$ \\
\hspace{0.16in}$>$ 125  & \hspace{0.12in}0 & \hspace{0.00in}$0.10 \pm 0.04$ & $\hspace{0.06in}0.034 \pm 0.034 $ & $0.057\pm 0.015$ & $4.66\pm 0.14$ &  $5.35\pm 0.11$ \\
\hline
\end{tabular}
\end{table*}

\section{Systematic uncertainties}
The GGM signal efficiency was determined from MC over an area of
GGM parameter space that ranges from $400$ to \unit[800]{GeV}
for the gluino mass, and from $\unit[50]{GeV}$ to within
$\unit[20]{GeV}$ of the gluino mass for the bino mass.
The efficiency increases smoothly
from $\unit[5]{\%}$ to $\unit[29]{\%}$ for $(\mass{\gluino}/\mass{\neutralino}) =
\unit[(400/50)]{GeV}$ to $\unit[(800/780)]{GeV}$; for low \mass{\neutralino},
the bulk of the \MET distribution lies below the cut of \unit[125]{GeV}, leading
to a less efficient selection.
The UED signal efficiency, also determined from MC, increases smoothly from
$\unit[37]{\%}$ for $1/R = \unit[700]{GeV}$ to $\unit[44]{\%}$ for $1/R =
\unit[1200]{GeV}$.

The various relative systematic uncertainties on the extraction of the
GGM and UED signal cross sections are summarised in
\Tab~\ref{tab:systematics} for the example GGM and UED samples,
including the \unit[3.4]{\%}
uncertainty on the integrated luminosity~\cite{Aad:2011dr,*lumi2010} and the
\unit[0.7]{\%} uncertainty on the trigger requirement. Some
sources of systematic uncertainty depend on kinematic properties and
are expected to differ between GGM and UED. The resulting ranges
provided in the following are ordered from low to high in the associated scales,
i.e. $\mass{\gluino},\mass{\neutralino}$ and $1/R$ in the case of GGM
and UED, respectively.

Uncertainties on the efficiency for reconstructing and identifying the
$\gamma\gamma$ pair arise from uncertainties on the impact of photon
quality and isolation cuts, the scale of the photon \ET cut, the
detailed material composition of the detector and from limitations in
the MC description of the photon identification variables.  The photon
selection efficiency in the MC was tuned to match the efficiency
observed in data by shifting the means of the discriminating variables
used in the photon selection, such as the shower widths. The resulting
estimate of the photon identification efficiency, tuned in this way,
was within \unit[5]{\%} of that of the un-tuned MC for all values of \ET and $\eta$.
A systematic uncertainty was assigned to the tuning
according to the uncertainty on the background contamination in the
data sample, as well as by replacing the data with a MC sample
generated with the inactive material in the region between the 
interaction point and calorimeter surface increased in a manner that
reflects the uncertainty on the actual material distribution. 
An additional systematic uncertainty on the overall photon selection
arises from the uncertainty on the overall amount of material in front
of the calorimeter, and has been estimated using the same MC
sample. The systematic uncertainty associated with removing photons
in problematic regions of the calorimeter was derived by comparing, in
MC and data, the fraction of photons within a fiducial region
surrounding areas of degraded detector response. Combined in quadrature, these
sources yielded a systematic uncertainty of \unit[3.6]{\%} for both GGM and
UED. 

The systematic uncertainty on the efficiency of the
isolation requirement was estimated by comparing distributions of the
isolation variable for identified photons, after 
subtracting the estimated contributions from jets mistakenly identified as photons,
with distributions of the isolation variable for MC photons. This yielded an
uncertainty between \unit[8.3]{\%} and \unit[0.5]{\%}~(GGM) and
of \unit[1.5]{\%}~(UED). The systematic uncertainty arising from 
the photon energy scale (known better than \unit[1.5]{\%}) was found to be negligible.
The influence of pileup, evaluated by comparing MC samples with
different pileup configurations, led to a systematic uncertainty of
\unit[1.0]{\%}. Systematic effects due to the 
$\MET$ reconstruction, estimated by varying the cluster energies
within established uncertainty ranges, and by varying the expected
$\MET$ resolution between the measured performance and MC
expectations, contributed an uncertainty of \unit[10.9]{\%} to
\unit[0.8]{\%}~(GGM) and \unit[2.1]{\%} to \unit[0.9]{\%}~(UED) on the
signal efficiency.
Finally, an uncertainty between \unit[3.8]{\%} and \unit[1.1]{\%} was
attributed to the limited statistics of the MC signal samples. Added
in quadrature, the total systematic uncertainty on the signal yield
varies between \unit[15.1]{\%} and \unit[5.6]{\%} (GGM) and
\unit[5.8]{\%} and \unit[5.4]{\%} (UED).  

\begin{figure}[b]
  \begin{center}
    \includegraphics[width=0.48\textwidth]{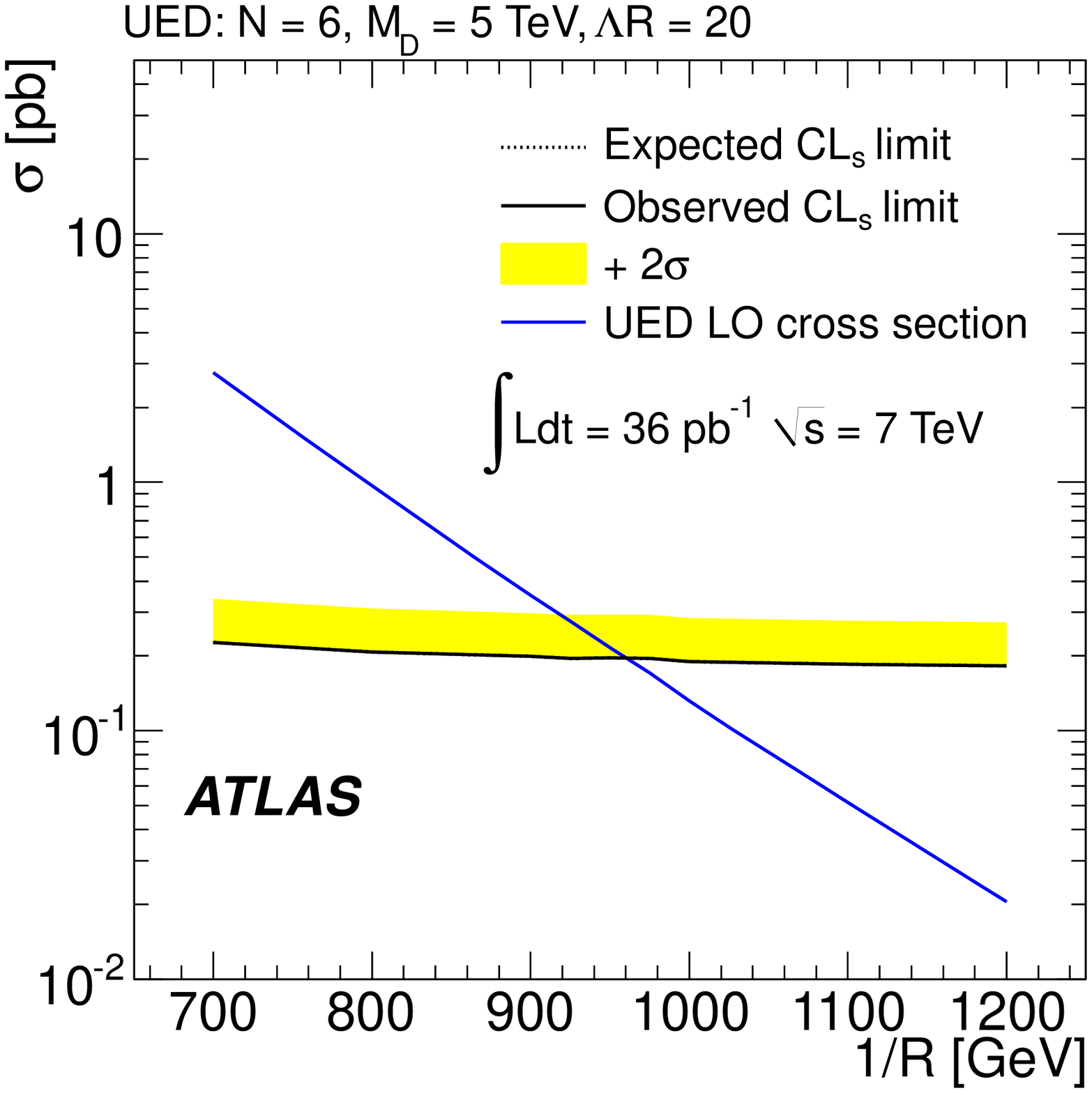}
  \end{center}
  \caption{Expected and observed \unit[95]{\%}~\CL upper limits on the UED production cross
    section, and the LO theory cross section prediction, as a function
    of $1/R$. The observed limit, the $\pm1\sigma$ expected error
    bands and the $-2\sigma$ expected error band are degenerate with the expected limit.
    The UED model parameters are $N=6$, $M_D = \unit[5]{TeV}$ and $\Lambda R = 20$. }
  \label{fig:ued:limit}
\end{figure}

In the case of GGM, the dominant theoretical uncertainties arise from
the limited knowledge of the proton PDF and the appropriate
factorisation and renormalisation scales.
The PDF uncertainties in the GGM cross sections were evaluated by using
the {\tt CTEQ6.6M} PDF error sets~\cite{Stump:2003yu} in the \Prospino
cross section calculation and are estimated to be between
\unit[12]{\%} and \unit[25]{\%}. 
The factorisation and renormalisation scales in the NLO
\Prospino calculation were increased and decreased by a factor of two,
leading to a systematic uncertainty between 
\unit[16]{\%} and \unit[19]{\%}.
The total theoretical uncertainty for GGM ranges from \unit[20]{\%} to
\unit[31]{\%}.
In the case of UED, NLO calculations of the cross sections are not yet
available, leading to arbitrarily large scale errors that artificially
degrade the derived limit. Thus, the LO cross sections were used with
no assignment of theoretical uncertainties, e.g. PDF or scale
uncertainties. The UED results are presented in a way that allows
limits to be derived for other assumptions about the size and
uncertainty of the UED cross section (see \Fig~\ref{fig:ued:limit}).

\begin{table}[b]
  \caption{Relative systematic uncertainties on the expected signal
    yield for GGM with $(\mass{\gluino}/\mass{\neutralino}) = 
    \unit[(600/300)]{GeV}$ and UED with $1/R = \unit[900]{GeV}$. No
    PDF and scale uncertainties are give for the UED case as the cross
    section is evaluated only to LO. More detail is provided in the text.}
  \label{tab:systematics}
  \center
\begin{tabular}{lrr} 
  \hline
  Source of uncertainty & \multicolumn{2}{c}{Uncertainty} \\
& \multicolumn{1}{c}{GGM} & \multicolumn{1}{c}{UED} \\ \hline
 Integrated luminosity           & \multicolumn{2}{c}{$\hspace{0.04in}3.4$\%} \\
 Trigger                         & \multicolumn{2}{c}{$\hspace{0.04in}0.7$\%} \\
 Photon identification           & \multicolumn{2}{c}{$\hspace{0.04in}3.6$\%} \\
 Photon isolation                & $\hspace{0.02in}3.4$\%  & $\hspace{0.02in}1.5$\% \\
 Effect of pileup                & \multicolumn{2}{c}{$\hspace{0.02in}1.0$\%} \\
 $\MET$ reconstruction and scale & $\hspace{0.02in}3.7$\% & $\hspace{0.02in}1.1$\% \\
 Signal MC statistics            & $\hspace{0.02in}2.6$\%  & $\hspace{0.02in}1.2$\% \\
 \hline
 Total signal uncertainty        & $\hspace{0.02in}7.6$\%&  $\hspace{0.02in}5.4$\% \\
 \hline
 PDF and scale                   & $\hspace{0.02in}25$\%& $\hspace{0.02in}-$  \\ 
 \hline
 Total                           & $\hspace{0.02in}26$\%&  $\hspace{0.02in}5.4$\% \\
 \hline
 \end{tabular}
\end{table}

\begin{figure}[t]
  \begin{center}
    \includegraphics[width=0.48\textwidth]{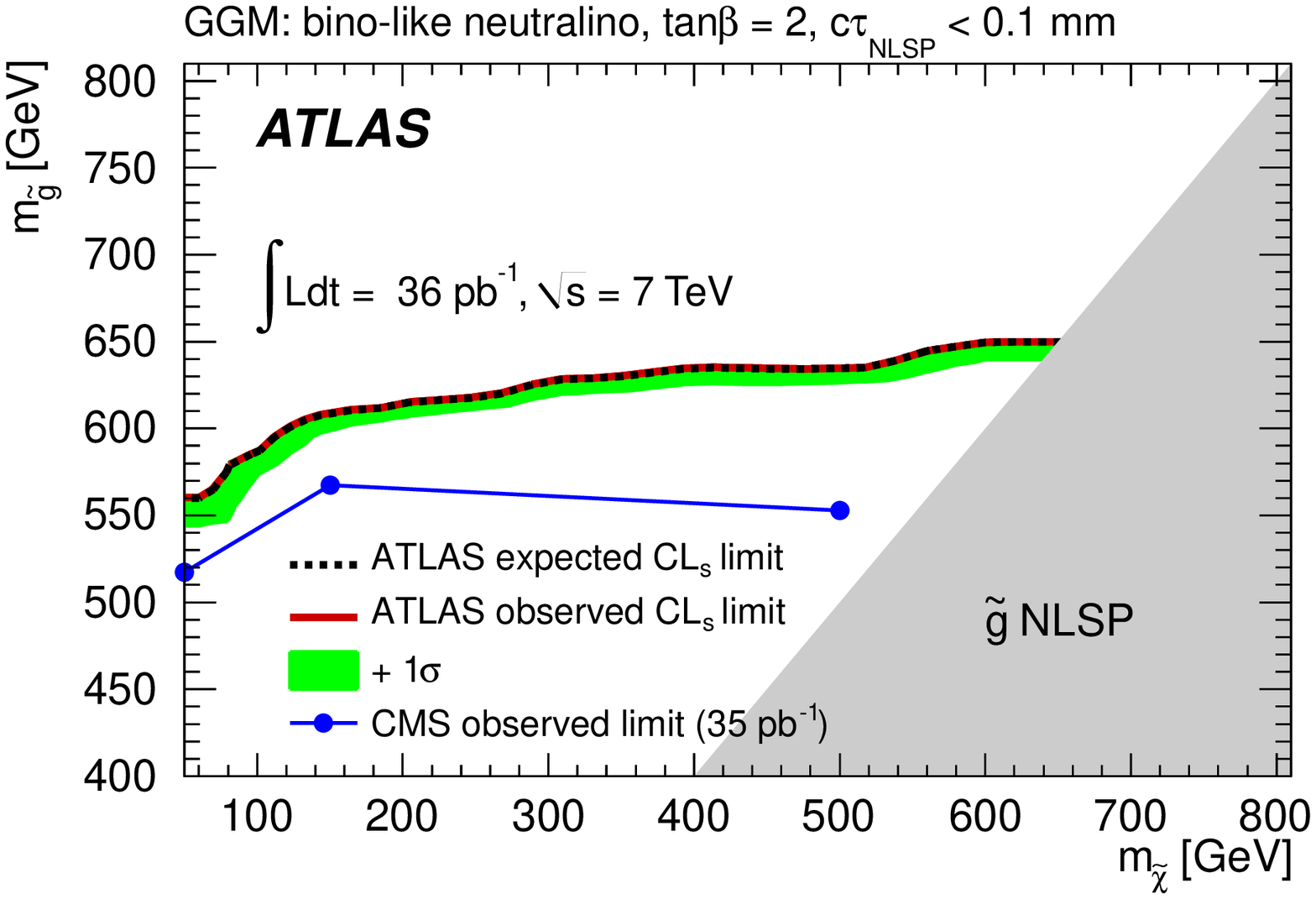}
  \end{center}
  \caption{Expected and observed \unit[95]{\%} \CL lower limits on the
    gluino mass as a function of the neutralino mass in the GGM model
    with a bino-like lightest neutralino as NLSP (the grey area
    indicates the region where the NLSP is the gluino, which was 
    not considered here). The other sparticle masses
    are fixed to $\sim\unit[1.5]{TeV}$. Further model parameters are
    $\tan\beta=2$ and $c\tau_{\mathrm{NLSP}} < \unit[0.1]{mm}$. 
    The observed limit and the $-1\sigma$ expected error band are degenerate with the expected limit.
    CMS lower limits are from \Ref~\protect\cite{Chatrchyan:2011wc}.
  }
  \label{fig:ggm:limit}
\end{figure}

\section{Extraction of limits}
Based on the observation of zero events with $\MET>\unit[125]{GeV}$ and the
expected background of \nbackground events, a \unit[95]{\%}~\CL upper limit was set on the
number of events from new physics in the signal region.

Frequentist limits were derived using a profile likelihood ratio
$\Lambda(s)$~\cite{Cowan:2010js}, with the likelihood function of the
fit given by $L({n}|s,{b},{\btheta}) = P_{\mathrm{S}} \times \,
C_{\mathrm{Syst}}$; ${n}$ represents the number of observed events in
data, $s$ the strength of the new physics signal under consideration,
${b}$ the expected background contribution, and ${\btheta}$ the
systematic uncertainties. The $P_{\mathrm{S}}$ function is a Poisson
probability distribution for the number of signal-region events and
$C_{\mathrm{Syst}}$ represents the constraints on systematic
uncertainties, which are treated as nuisance parameters with a
Gaussian probability density.  The one-sided exclusion $p$-values were
obtained using the test statistic $\Lambda(s)$ distribution from
pseu\-do-experiments, and the $CL_s$ method was used to exclude possible
contributions from the signal~\cite{Read:2002hq}.

The number of events in the signal region from any scenario of physics
beyond the SM (BSM) was found to be less than \Nlimit at
\unit[95]{\%}~\CL. This number corresponds to \unit[95]{\%}~\CL upper
limits on production cross sections of \sigmaGGM in 
the GGM model (\mass{\neutralino}=\unit[150]{GeV}, $\mass{\gluino}=400-\unit[800]{GeV}$) and \sigmaUED in
the UED model ($1/R=700-\unit[1200]{GeV}$), shown as a function of $1/R$ for the case of
the UED model in \Fig~\ref{fig:ued:limit}. 
These upper limits on cross section include systematic
uncertainties on the background estimation, event selection and the
luminosity.

These results can be interpreted in terms of \unit[95]{\%}~\CL
exclusion limits on specific parameters of the two new physics models
considered.
Figure~\ref{fig:ued:limit} depicts the lower limit on the
curvature parameter $1/R$ in the context of the UED model
considered. The observed (expected) \unit[95]{\%}~\CL exclusion region is
\UEDlimit (\UEDlimitExp).
Figure~\ref{fig:ggm:limit} shows the expected and observed lower limit on the
GGM gluino mass as a function of the neutralino mass,
with all other sparticle masses, e.g. squarks,
set to $\sim\unit[1.5]{TeV}$.
In addition to the experimental systematic
uncertainties, the limit also takes into account theoretical uncertainty
on the production cross section.
The limit depends only weakly on the neutralino mass, and a lower observed (expected) 
gluino mass limit of
\GGMlimit (\GGMlimitExp) is obtained for neutralino masses above \unit[50]{GeV}. For
comparison the lower limits from CMS~\cite{Chatrchyan:2011wc} on the gluino mass for neutralino
masses of $50$, $175$ and $\unit[500]{GeV}$ are
shown in the same figure.

The limits presented here, obtained with all squark
masses set to $\sim\unit[1.5]{TeV}$, are conservative, since values of
other strongly-charged sparticle (squark) masses that lie close to
the excluded gluino mass
increase the cross section for pair 
production of coloured SUSY particles, leading to a more stringent bounds on the gluino mass.
The impact of variations in the systematic uncertainty is small: the
observed limits change by less than \unit[2]{GeV} if the magnitude of
the systematic uncertainty is allowed to approach $0$.

\section{Conclusions}
A search for $\gamma\gamma$ events with large $\MET$,
conducted using a \integLumi~sample of \unit[7]{TeV} $pp$ collision data
recorded with the ATLAS detector at the LHC, found no evidence of an
excess above the SM expectation:
zero events were observed with an expected background of \nbackground.
The results were used to set a model-indepen\-dent \unit[95]{\%}~\CL
upper limit of \Nlimit events 
on the observed number of diphoton events from new physics in the 
region $\MET>\unit[125]{GeV}$.
\unit[95]{\%}~\CL upper limits were also set on the production cross section for two
particular models of
new physics: \sigmaGGM for
the GGM model (\mass{\neutralino}=\unit[150]{GeV}) and \sigmaUED for
the UED model. These limits are interpreted in terms of constraints on
parameters of specific new physics models. 
Under the GGM hypothesis, a lower limit on the gluino mass of \GGMlimit is determined for
bino masses above \unit[50]{GeV},
while a lower limit of \UEDlimitlow is set on the UED
compactification radius $R$. These limits provide the most stringent
tests of these models to date.

\begin{acknowledgement}
  \section*{Acknowledgements}
  
We thank CERN for the very successful operation of the LHC, as well as the
support staff from our institutions without whom ATLAS could not be
operated efficiently.

We acknowledge the support of ANPCyT, Argentina; YerPhI, Armenia; ARC,
Australia; BMWF, Austria; ANAS, Azerbaijan; SSTC, Belarus; CNPq and FAPESP,
Brazil; NSERC, NRC and CFI, Canada; CERN; CONICYT, Chile; CAS, MOST and
NSFC, China; COLCIENCIAS, Colombia; MSMT CR, MPO CR and VSC CR, Czech
Republic; DNRF, DNSRC and Lundbeck Foundation, Denmark; ARTEMIS, European
Union; IN2P3-CNRS, CEA-DSM/IRFU, France; GNAS, Georgia; \linebreak BMBF, DFG, HGF, MPG
and AvH Foundation, Germany; GSRT, Greece; ISF, MINERVA, GIF, DIP and
Benoziyo Center, Israel; INFN, Italy; MEXT and JSPS, Japan; CNRST, Morocco;
FOM and NWO, Netherlands; RCN, Norway; MNiSW, Poland; GRICES and FCT,
Portugal; MERYS (MECTS), Romania; MES of Russia and ROSATOM, Russian
Federation; JINR; MSTD, Serbia; MSSR, Slovakia; ARRS and MVZT, Slovenia;
DST/NRF, South Africa; MICINN, Spain; SRC and Wallenberg Foundation,
Sweden; SER, SNSF and Cantons of Bern and Geneva, Switzerland; NSC, Taiwan;
TAEK, Turkey; STFC, the Royal Society and Leverhulme Trust, United Kingdom;
DOE and NSF, United States of America.

The crucial computing support from all WLCG partners is acknowledged
gratefully, in particular from CERN and the ATLAS Tier-1 facilities at
TRIUMF (Canada), NDGF (Denmark, Norway, Sweden), CC-IN2P3 (France),
KIT/GridKA (Germany), INFN-CNAF (Italy), NL-T1 (Netherlands), PIC (Spain),
ASGC (Taiwan), RAL (UK) and BNL (USA) and in the Tier-2 facilities
worldwide.

\end{acknowledgement}

\bibliographystyle{atlasnote}
\bibliography{DiPhotonMetPRL}

\clearpage
\hugehead
\begin{flushleft}
{\Large The ATLAS Collaboration}

\bigskip

G.~Aad$^{\rm 48}$,
B.~Abbott$^{\rm 111}$,
J.~Abdallah$^{\rm 11}$,
A.A.~Abdelalim$^{\rm 49}$,
A.~Abdesselam$^{\rm 118}$,
O.~Abdinov$^{\rm 10}$,
B.~Abi$^{\rm 112}$,
M.~Abolins$^{\rm 88}$,
H.~Abramowicz$^{\rm 153}$,
H.~Abreu$^{\rm 115}$
% ...
\bigskip

% ...
$^{10}$ Institute of Physics, Azerbaijan Academy of Sciences, Baku, Azerbaijan\\
$^{11}$ Institut de F\'isica d'Altes Energies and Universitat Aut\`onoma  de Barcelona and ICREA, Barcelona, Spain\\
% ...
$^{48}$ Fakult\"{a}t f\"{u}r Mathematik und Physik, Albert-Ludwigs-Universit\"{a}t, Freiburg i.Br., Germany\\
$^{49}$ Section de Physique, Universit\'e de Gen\`eve, Geneva, Switzerland\\
% ...
$^{88}$ Department of Physics and Astronomy, Michigan State University, East Lansing MI, United States of America\\
% ...
$^{111}$ Homer L. Dodge Department of Physics and Astronomy, University of Oklahoma, Norman OK, United States of America\\
$^{112}$ Department of Physics, Oklahoma State University, Stillwater OK, United States of America\\
% ...
$^{115}$ LAL, Univ. Paris-Sud and CNRS/IN2P3, Orsay, France\\
% ...
$^{118}$ Department of Physics, Oxford University, Oxford, United Kingdom\\
% ...
$^{153}$ Raymond and Beverly Sackler School of Physics and Astronomy, Tel Aviv University, Tel Aviv, Israel\\
% ...
\end{flushleft}


\end{document}